\documentclass[sigconf,natbib=true,backend=bibtex, review=False]{acmart}

\AtBeginDocument{%
  }

\usepackage{graphicx}
\usepackage{adjustbox}
\usepackage{booktabs}
\usepackage{multicol}
\usepackage{multirow}
\usepackage{subcaption}
\usepackage{balance}

\newcommand{\bsacmplace}{ACM, New York, NY, USA}

\makeatletter
\def\@mkbibcitation{\bgroup
\let\@vspace\@vspace@orig
\let\@vspacer\@vspacer@orig
\def\@pages@word{\ifnum\getrefnumber{TotPages}=1\relax page\else pages\fi}%
\def\footnotemark{}%
\def\\{\unskip{} \ignorespaces}%
\def\footnote{\ClassError{\@classname}{Please do not use footnotes
    inside a \string\title{} or \string\author{} command! Use
    \string\titlenote{} or \string\authornote{} instead!}}%
\def\@article@string{\ifx\@acmArticle\@empty{\ }\else,
    Article~\@acmArticle\ \fi}%
\par\medskip\small\noindent{\bfseries ACM Reference Format:}\par\nobreak
\noindent\bgroup
\def\\{\unskip{}, \ignorespaces}\authors\egroup. \@acmYear. \@title
\ifx\@subtitle\@empty. \else: \@subtitle. \fi
\if@ACM@nonacm\else
  \if@ACM@journal@bibstrip
    \textit{\@journalNameShort}
    \@acmVolume, \@acmNumber \@article@string (\@acmPubDate),
    \ref{TotPages}~\@pages@word.
  \else
    In \textit{\@acmBooktitle}%
    \ifx\@acmEditors\@empty\textit{.}\else
      \andify\@acmEditors\textit{, }\@acmEditors~\@editorsAbbrev.%
    \fi\
    \bsacmplace%
    \@article@string\unskip, \ref{TotPages}~\@pages@word.
  \fi
\fi
\ifx\@acmDOI\@empty\else\@formatdoi{\@acmDOI}\fi
\par\egroup}
\makeatother

\usepackage{cleveref}
\newboolean{showcomments}
\setboolean{showcomments}{false} %
\ifthenelse{\boolean{showcomments}}
{
	\newcommand{\nb}[3]{
		{\colorbox{#2}{\bfseries\sffamily\scriptsize\textcolor{white}{#1}}}
		{\textcolor{#2}{$\blacktriangleright$\textsf\small{#3}$\blacktriangleleft$}}}
	 
}{
	\newcommand{\nb}[3]{}
} 

\usepackage{enumitem} 
\newcommand{\uls}{\begin{itemize}[leftmargin=*]}
\newcommand{\ule}{\end{itemize}}
\newcommand{\ols}{\begin{enumerate}[leftmargin=*]}
\newcommand{\ole}{\end{enumerate}}

\newcommand{\para}[1]{\paragraph{\textnormal{\textbf{#1}}}}

\begin{document}

\title{Towards a Relevance Posterior in Neural Information Access}

\settopmatter{authorsperrow=4}

 \author{Andrew Parry}
\affiliation{\institution{University of Glasgow}
  \city{Glasgow} \country{UK}}

 \author{Emmanouil Georgios Lionis}
\affiliation{\institution{University of Glasgow}
  \city{Glasgow} \country{UK}}

 \author{Debasis Ganguly}
\affiliation{\institution{University of Glasgow}
  \city{Glasgow} \country{UK}}

 \author{Sean MacAvaney}
\affiliation{\institution{University of Glasgow}
  \city{Glasgow} \country{UK}}

\copyrightyear{2026}
\acmYear{2026}
\setcopyright{cc}
\setcctype{by}
\acmConference[SIGIR '26] {Proceedings of the 49th International ACM SIGIR Conference on Research and Development in Information Retrieval}{July 20--24, 2026}{Melbourne, VIC, Australia.}
\acmBooktitle{Proceedings of the 49th International ACM SIGIR Conference on Research and Development in Information Retrieval (SIGIR '26), July 20--24, 2026, Melbourne, VIC, Australia}
\acmISBN{979-8-4007-2599-9/2026/07}
\acmDOI{10.1145/3805712.3808541}
\renewcommand{\shortauthors}{Andrew Parry, Emmanouil Georgios Lionis, Debasis Ganguly, \& Sean MacAvaney}

\begin{abstract}
Modern information retrieval systems typically operationalise relevance as a query-conditional score computed at inference time. This design choice has become dominant such that alternative decompositions of relevance are rarely discussed, despite the long history of document and query priors in probabilistic retrieval and large-scale search. As neural ranking models grow more computationally expensive and retrieval pipelines expand to include multi-stage ranking, recommendation, and retrieval-augmented generation, this monolithic view of query-time scoring becomes increasingly limiting. We argue that modern information access systems are more naturally understood as performing approximate posterior inference, in which relevance is refined through a staged combination of query-dependent likelihoods and query-independent priors. We extend classical probabilistic retrieval formalisms to contemporary learned systems and show how explicit likelihood–prior decomposition exposes new opportunities to shift computation offline while disentangling document-level and interaction-level beliefs. We present empirical evidence that incorporating query-independent document utility can complement existing rankers and improve effectiveness with minimal query-time computation (solely score fusion). Concretely, a learned prior improves first-stage retrieval through rank fusion (up to $\Delta$ nDCG@10 $\approx 0.046$ on TREC DL-2019 and $\approx 0.029$ on TREC DL-2020) and also improves downstream re-ranking, with the largest gains observed for the LLM re-ranker RankZephyr (up to $\Delta$ nDCG@10 $\approx 0.054$ on TREC DL-2020). Finally, we discuss how this decomposition connects to broader information access and outline research directions for designing retrieval systems that explicitly allocate modelling capacity between offline priors and online interaction.

\end{abstract}



\maketitle

\section{Introduction}

Relevance in information retrieval reflects the degree to which a document satisfies an information need~\cite{saracevic:2007}. This quantity is not directly observed~\citep{belkin:1980, mizzaro:1997}, but inferred from partial evidence about the query, the document, and the retrieval context~\cite{robertson:1976}. In modern neural retrieval systems, this inference is operationalised as a single scalar score computed at query time for each candidate document~\citep{nogueira:2019, karpukhin:2020}, implicitly committing systems to resolving nearly all relevance-related uncertainty during query--document interaction under strict latency constraints~\cite{dean:2009}. Yet some beliefs about documents are stable across queries~\cite{azzopardi:2008, chang:2024} and can be formed in advance and reused across interactions~\cite{peng:2008}.

\begin{figure}
    \centering
    \includegraphics[width=\linewidth]{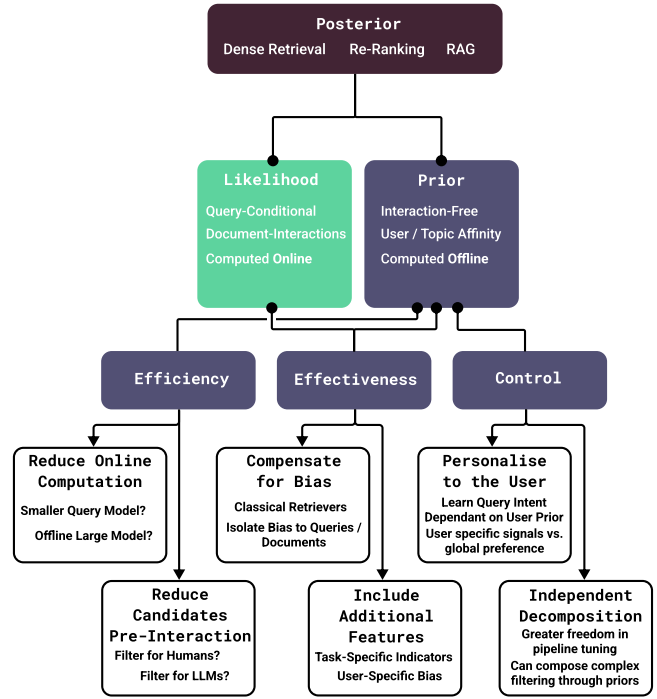}
    \caption{An illustration of prior-likelihood decomposition and its downstream benefits with respect to three pillars: Efficiency, Effectiveness and Control. Our empirical investigation targets both efficiency and effectiveness by reusing a static pruning model to improve ranking precision.}
    \Description{A flowchart showing a posterior box at the top containing dense retrieval, re-ranking, and RAG. It decomposes into two boxes: likelihood (query-conditional, document interactions, computed online) and prior (interaction-free, user/topic affinity, computed offline). Below, three pillars show downstream benefits: efficiency (reduce online computation, reduce candidates), effectiveness (compensate for bias, include additional features), and control (personalise to user, independent decomposition).}
    \label{fig:main}
\end{figure}

\paragraph{Perspective.}
We treat retrieval as approximate inference over a relevance posterior. Under this view, modern pipelines implement \emph{staged posterior refinement}: (i) offline, query-independent components form reusable document-side beliefs (prior terms)~\cite{robertson:1995, peng:2008, azzopardi:2008}, and (ii) online, query-conditioned components contribute evidence that resolves intent (likelihood-like interactions). This refinement is mediated by approximation operators that trade computation for posterior fidelity, including candidate truncation~\cite{nogueira:2019}, approximate nearest-neighbour recall~\cite{malkov:2020, johnson:2021}, and bounded context budgets in retrieval-augmented generation~\cite{lewis:2020, guu:2020}. This lens clarifies which features can be amortised offline, which must remain online, and how they should be composed within deployment budgets. We illustrate the benefits of this formulation in Figure \ref{fig:main}.  

Classical probabilistic retrieval models explicitly distinguish between query-dependent evidence and query-independent document effects, often via document priors~\cite{robertson:1995, peng:2008}. Importantly, alternative decompositions of relevance are possible without altering the underlying ranking objective, leveraging both query--document interactions and features of documents computed offline. The neural setting alters the practical character of this distinction in two related ways. First, powerful pretrained language models can serve as offline document scorers, running once at index time and caching their output for reuse across all queries at a cost that is amortised over the full query workload. Whereas classical document priors were largely restricted to corpus statistics and link-graph signals, learned neural priors can capture semantic quality, authority, and utility at a representational depth that heuristic signals cannot reach, while remaining strictly query-independent by construction. Second, this offline capacity is rarely exploited in contemporary pipelines. Instead, end-to-end gradient-based training tends to absorb stable document-side regularities into the same parameters that resolve query intent, because both signals are present in relevance supervision and neither is architecturally isolated~\cite{braga:2025}. Feature-based learning-to-rank exhibits a related entanglement, but the neural case is sharper in degree: representations are shared across query and document sides, supervision is end-to-end, and the resulting model cannot be decomposed into reusable offline and online components without retraining. The combination of these factors, namely the availability of strong offline scorers and the tendency of online models to recover document-side structure that need not be re-estimated per query, motivates treating the prior--likelihood split as an explicit design variable rather than an artefact of inference.

Neural rankers do not model query--document interactions in isolation. In practice, they jointly model query-specific signals and corpus-level regularities arising from training data distributions, negative sampling procedures~\cite{karpukhin:2020}, and document properties that correlate with relevance across many queries such as readability or general utility~\cite{zamani:2018, geva:2021}. This phenomenon has been isolated, explicitly showing that much of the optimisation effort within a neural model is in fact learning characteristics of a relevant document \emph{a priori} of a particular query~\cite{braga:2025}, even when the optimisation objective does not explicitly consider a prior over documents. As a result, evidence of relevance that can be captured offline is conflated with evidence that must be resolved at interaction time. Crucially, this entanglement incurs an efficiency cost: modelling effort and computation are repeatedly expended at query time to recover structure that could be learned once offline and reused across interactions.

Query-independent document utility (quality, authority, or expected utility marginalised over the query distribution~\cite{peng:2008, chang:2024}) provides a concrete instance of such offline modelling. Because these properties do not depend on a specific query~\cite{peng:2008}, they can be learned and cached entirely offline. However, document utility is only one instance of a broader class of priors that may also condition on users, domains, temporal context, or system-level constraints.

Within this perspective, our contributions are threefold. \textbf{1}) We frame and advocate for modern learned information access systems using an explicit prior--likelihood decomposition, providing a concrete language for reasoning about how relevance estimation is distributed across offline and online components. \textbf{2}) We show empirically that learning a query-independent prior offline can compensate for biases and limitations in initial retrieval without increasing query-time cost. \textbf{3}) We outline future research directions on how offline priors and query-time models should be composed to balance effectiveness, efficiency, and control under modern deployment constraints.

\section{Related Work}
Neural information retrieval has largely centred on query-conditional ranking. Whether through cross-encoder re-ranking~\cite{macavaney:2019, nogueira:2019}, dual-encoder retrieval~\cite{karpukhin:2020}, or learned sparse representations~\cite{formal:2021}, a common pattern persists: a lightweight retriever generates candidates that are progressively refined by increasingly expensive interaction~\cite{pradeep:2021}. Improvements in effectiveness have coincided with increased interaction complexity~\cite{khattab:2020} and higher serving cost, and while distillation reduces online burden~\cite{hofstatter:2020, lin:2020}, reranker capacity continues to grow~\cite{pradeep:2023, parry:2025}. Some architectures shift computation offline: bi-encoders precompute whole or per-token document representations~\cite{karpukhin:2020, khattab:2020}, and document expansion generates surrogate queries offline~\cite{nogueira:2019b}, but these strategies target the \emph{likelihood} (improving query--document matching) rather than leveraging this offline computation to model document utility independently.

\para{Learning to Rank.}
Learning-to-rank (LTR) methods combine heterogeneous document features, including query-independent signals such as PageRank~\cite{page:1999}, document length~\cite{robertson:1995}, and spam scores~\cite{cormack:2010}, within a supervised framework~\cite{liu:2009}. LTR is therefore the most direct precedent for incorporating static document signals into ranking. However, LTR treats all features as inputs to a single monolithic scorer trained end-to-end; the distinction between query-dependent and query-independent evidence is not preserved in the learned model and cannot be exploited at deployment time. Once trained, an LTR model requires all features at inference, and updating a single signal (e.g., a freshness estimate) necessitates retraining or recalibrating the full model. By contrast, the prior--likelihood decomposition maintains query-independent signals as separable, reusable components that can be updated, audited, or replaced without modifying the query-time model. The distinction is architectural rather than representational: both approaches can, in principle, access the same information, but the decomposition exposes modularity that monolithic scoring obscures. Critically, this modularity is operationally meaningful only in the neural setting: classical LTR feature pipelines required all signals at serving time regardless, whereas a cached neural prior contributes to scoring at zero marginal document-side cost.

\para{Priors in Information Retrieval.}
Query-independent document signals have long played a central role in large-scale search. PageRank~\cite{page:1999} and HITS~\cite{kleinberg:1999} established that link-graph authority, computed entirely offline, substantially improves retrieval when combined with content-based matching. Document quality estimation~\cite{chang:2024}, spam detection~\cite{cormack:2010}, and readability scoring have similarly been computed offline and incorporated as document priors~\cite{peng:2008}. These signals shape which documents are indexed, retrieved, or promoted, providing a reusable layer of corpus-level modelling independent of individual queries.

Recent work has revisited static document modelling in neural IR, particularly for static corpus filtering~\cite{cormack:2007, chang:2024}. Neural quality estimators can remove large fractions of a corpus while preserving downstream effectiveness, reducing indexing cost, storage, and latency. Crucially, these models run entirely offline and can leverage powerful architectures (including large language models) without affecting query-time performance~\cite{chang:2024}.

However, current practice largely treats static scores as binary gating decisions applied during indexing~\cite{chang:2024}. Once pruning is complete, the learned document signal is discarded and does not participate in ranking. This creates a disconnect between offline document modelling and online ranking, despite both targeting the same relevance objective. A key distinction of our perspective is \emph{lifecycle and composition}: rather than treating offline scores as gates (index inclusion) or as additional features inside a monolithic ranker, we treat them as first-class priors that persist throughout the pipeline. 

\para{Topic Models and Latent Structure.}
Latent variable models have long separated corpus-level structure from query-specific evidence. Probabilistic latent semantic analysis~\cite{hofmann:1999} and latent Dirichlet allocation~\cite{blei:2003, wei:2006} model documents as mixtures of topics estimated offline and reused across queries, capturing corpus-wide regularities that would otherwise be rediscovered at query time. This convention aligns with the view of retrieval as an interaction between an information need and a latent information space~\cite{belkin:1980}, in which queries are partial expressions of an underlying need and latent variables capture structure not directly observable in the query. Modern neural systems rarely use topic models directly but rely on the same principle: relevance is mediated by latent corpus structure learnable offline. Neural representations absorb this structure implicitly through pretraining, but the separation between corpus-level regularities and query-specific evidence remains fundamental. Our work extends these notions by arguing that systems can benefit from reintroducing explicit document-side components.

\para{Retrieval-Augmented Generation.}
RAG systems~\cite{lewis:2020} introduce a qualitatively different approximation boundary: rather than ranking over candidates, the system selects evidence under a hard context budget. Recent work has shown that retrieval quality directly governs generation quality~\cite{shi:2023}, and that LLM generators are sensitive to document ordering, redundancy, and noise in the retrieved context~\cite{liu:2024}. These findings suggest that document-level properties such as reliability, conciseness, and answerability are primary factors affecting relevance in RAG \cite{DBLP:conf/ecir/TianGM25}, yet are rarely explicitly modelled. Under a prior likelihood--decomposition, such properties are naturally represented as prior terms that can filter or reweight candidates before context construction \cite{DBLP:conf/ecir/ChandraGO26}, reducing the burden on the generator to identify useful evidence from noisy retrieval.

\section{Bridging Probabilistic IR to Learned Systems}
\label{sec:bridge}
We first outline the connection between classical probabilistic term modelling and how the formal language of those models can be helpful in reasoning over modern learned systems.

\para{Classical Probabilistic Retrieval}
We consider ad-hoc retrieval over a corpus $\mathcal{C}$. Classical probabilistic retrieval models rank documents ($d$) given a query ($q$) according to posterior probability of relevance, following the Probability Ranking Principle (PRP)~\cite{robertson:1977}. In language-model-based formulations, this is expressed through a query-likelihood decomposition:
\begin{equation}
p(d \mid q) = \frac{p(q \mid d)\,p(d)}{p(q)},
\label{eq:bayes}
\end{equation}
where the constant marginal $p(q) = \sum_{d \in \mathcal{C}} p(q \mid d) \ p(d)$ need not be explicitly computed as it does not affect the relative ordering of the query-conditional document estimates.

In Equation~\ref{eq:bayes}, $p(d)$ is a query-independent document prior and $p(q \mid d)$ denotes a query-dependent likelihood. Under a multinomial unigram language model with the term independence assumption~\cite{hiemstra2001using, ponte:2008}, the query likelihood factorises over query terms and is typically smoothed against the collection:
\begin{align}
P(q \mid d) &= \prod_{t \in q} P(t \mid d) \\
            &= \prod_{t \in q} \left( \lambda P_{MLE}(t \mid d) + (1-\lambda) P_{MLE}(t \mid \mathcal{C}) \right) \\
            &= \prod_{t \in q} \left( \lambda \frac{f(t,d)}{f(\cdot,d)} + (1-\lambda) \frac{f(t,\mathcal{C})}{f(\cdot)} \right),
\end{align}
where (a) $P_{MLE}(t \mid d)$ and $P_{MLE}(t \mid \mathcal{C})$ respectively denote the maximum likelihood estimates of sampling $t$ either from the document $d$ or from the collection, (b) $f(\cdot, d) = |d| = \sum_{w \in d} f(w, d)$ represents the document length and $f(\cdot)$ denotes the cardinality of the vocabulary multiset (i.e.\ the collection size), and (c) $\lambda$ controls the mixture weight on the document-conditional sampling probability relative to the background collection model.

Historically, this factorisation was rarely treated as a modelling decision to be optimised. The likelihood received primary attention~\cite{robertson:1995}, while the prior was instantiated through heuristics: document length normalisation~\cite{robertson:1995}, collection statistics~\cite{ponte:2008}, or uniformity assumptions~\cite{zhai2004study}. This parallels latent-variable models in IR: probabilistic latent semantic analysis~\cite{hofmann:1999,ganguly:2015} and latent Dirichlet allocation~\cite{blei:2003, wei:2006, DBLP:conf/sigir/GangulyLJ13}, where corpus-level structure is estimated offline and reused across queries. The separation between corpus-level regularities and query-specific evidence has always been present; what remained unexplored was how these components should be learned, composed, and reused.

Modern neural IR systems collapse this decomposition into a single query-conditioned scoring function. Whether implemented as bi-encoder similarity, cross-encoder re-ranking, or learned sparse retrieval, contemporary models absorb corpus-level effects into parameters through training data, negative sampling, and optimisation dynamics. The document prior has not disappeared; it has been marginalised into the learned representation space, thereby implying a change in inference strategy rather than in the learning objective. This marginalisation is a consequence of joint optimisation: because query-independent and query-dependent signals are both present in relevance supervision, gradient flow entangles them within shared parameters in a way that classical term-weighting models, which never shared parameters across these two roles, did not. The practical implication is that structure which could be estimated once offline must instead be recovered at query time, repeatedly, across every interaction.

\para{Conditioning as a Design Variable}
Conditioning choices determine not only which information sources are used but \emph{when} they enter inference. Writing relevance as a marginalisation over latent factors $z$ (corpus structure~\cite{robertson:1976}, user populations~\cite{he:2017}, task constraints or similar semantics~\cite{ganguly:2015}),
\begin{equation}
  p(d \mid q) = \sum_{z} p(d \mid q, z)\, p(z \mid q),
  \label{eq:latent-marginal}
\end{equation}
practical systems must choose a subset $x \subset z$ to condition on explicitly, marginalising the rest. Conditioning on $x$ yields
\begin{equation}
  p(d \mid q, x) \;\propto\; p(q \mid d, x)\, p(d \mid x),
  \label{eq:conditioned-posterior}
\end{equation}
committing the system to resolving $x$ at inference time. Each commitment carries a cost. The central design question is therefore \emph{which variables to retain at each pipeline stage} given a fixed \textit{computational budget}. Variables whose influence on document ordering is stable across queries (quality, authority, domain suitability) are natural candidates for offline resolution via priors; variables whose influence is query-specific (topical match, facet selection) must remain online. In neural systems this distinction is particularly consequential: a bi-encoder document embedding, for instance, already amortises document-side computation offline, but the training objective entangles prior-like corpus utility with likelihood-like query-predictive signal within the same representation, preventing either from being independently updated, audited, or reused across pipelines.

Conditioning on a variable $x$ that is not document-dependent preserves the ranking because:
\begin{equation}
  \text{if } x \perp\!\!\!\perp d \mid q, \quad
  \arg\max_{d}\, p(d \mid q)
    = \arg\max_{d}\, p(d \mid q, x).
  \label{eq:cond-invariance}
\end{equation}
This holds whenever $x$ is a property of the query, session, or retrieval context that does not induce a document-dependent score transformation (e.g., latency budget, user locale, task type). By contrast, conditioning on a document-dependent variable \emph{does} change the ranking, precisely the mechanism exploited by priors.

\para{Priors as Amortised Inference}

\emph{Amortised inference} refers to the strategy of performing expensive computation once offline and reusing the result across many future instances, rather than repeating the computation each time. PageRank~\cite{page:1999}, which estimates document authority once and reuses it across all queries, is a canonical example of amortised inference in retrieval. The neural setting extends this principle qualitatively: whereas PageRank is a graph statistic, a neural quality estimator can capture semantic utility, structural completeness, and domain reliability at depth that heuristic signals cannot reach, while remaining strictly query-independent by construction and cacheable at index time.

Document priors encode structure that is invariant across queries, estimated once offline and reused at query time across all query--document scoring interactions. The decomposition in Equation~\ref{eq:bayes} thus trades offline computation for reduced query-time cost, reframing priors not as heuristic components but as reusable outputs of amortised inference, quantities whose estimation has been paid for in advance.

In multi-stage retrieval systems, marginalisation is temporary. Early stages aggressively marginalise relevance factors under tight latency budgets. For example, a first-stage retriever such as BM25 or a bi-encoder effectively marginalises over user intent, document quality, and contextual factors, scoring documents using only surface lexical or semantic match. As the candidate set shrinks, later stages can afford to condition on richer evidence variables $x_1, \ldots, x_k$, such as cross-attention features between query and document tokens, user interaction history, or document quality signals, progressively refining the relevance estimate:
\[
p^{(0)}(d) \rightarrow p^{(1)}(d \mid q, x_1) \rightarrow \cdots \rightarrow p^{(k)}(d \mid q, x_{1:k}).
\]

More precisely, at each stage $i$ a new evidence factor $\ell_i(d; q)$ refines the current belief:
\begin{equation}
  p^{(i)}(d \mid q)
    \;\propto\;
    p^{(i-1)}(d \mid q)\;\ell_i(d;\, q),
  \label{eq:staged-update}
\end{equation}
where each $\ell_i$ may be query-dependent (e.g.\ a cross-encoder re-ranking score) or query-independent (e.g.\ a document quality or freshness filter). This staged update is analogous to message-passing in factor graphs~\cite{kschischang:2001}: just as each variable node in a factor graph aggregates messages from its neighbouring factors, here each retrieval stage contributes a local scoring factor, and the final ranking emerges from their product: the MAP configuration of the approximate posterior. The decomposition makes the \emph{order} and \emph{granularity} of evidence introduction an explicit design variable.

A similar pattern appears in agentic retrieval-augmented generation (RAG) systems~\cite{li:2025, jin:2025}, where an agent retrieves an initial set of documents, reasons over them, and then issues refined queries to retrieve again. Each retrieval--reasoning cycle can be understood as introducing a new evidence factor $\ell_i$ conditioned on the context accumulated so far, progressively sharpening the system's belief about which documents are relevant.

Over-aggressive marginalisation of stable factors introduces systematic inefficiency. When query-invariant structure is not retained, query-time models must repeatedly recover it. A neural ranker trained end-to-end, for instance, must implicitly learn to account for document length, authority, and recency (factors that could have been estimated once and supplied as priors), wasting model capacity on recovering structure that is stable across queries. Formally, such a model implicitly approximates
\begin{equation}
p(q \mid d;\theta) \approx \sum_x p(q \mid d, x)\,p(x \mid d)
\end{equation} 
within a single estimator. This entanglement obscures sources of variation, wastes computation on invariant structure, and reduces reusability across systems and tasks.

Retrieval systems are therefore more naturally understood as inference procedures, differing not in objective but in which variables are retained, when they are introduced, and under what constraints they are resolved.

\section{Systems as Approximate Inference}

Under the likelihood–prior framework introduced earlier, query--document interaction corresponds to the likelihood term: the component of relevance that cannot be estimated without observing the query~\cite{robertson:1995, lafferty:2001}. Many factors that influence relevance, such as general document quality, topical coverage, or retrievability, can be marginalised over queries and estimated offline~\cite{azzopardi:2008, chang:2024}. Query--document interaction therefore resolves residual uncertainty—those distinctions that depend on the specific information need being expressed. This framing grounds interaction in theory rather than architecture, linking it directly to what remains unresolved after offline document modelling.

\para{Query--Conditional Models}

In principle, query-conditional models are well-suited to resolving intent: determining which facets of information are relevant for a given query and how document evidence should be interpreted in that context~\cite{belkin:1980}. In practice, however, contemporary query models often serve a broader role, implicitly compensating for weak or missing document-side signals. From a decomposition perspective, intent resolution can be seen as the idealised role of query interaction once document utility is well estimated offline. We make this systems-level interpretation explicit via a compact taxonomy (Table~\ref{tab:systems_taxonomy}), which summarises where different pipelines encode query-independent beliefs, how they apply query-conditioned evidence, and which approximation operator (e.g., truncation or budgeting) ultimately constrains the reachable posterior mass.

Table \ref{tab:systems_taxonomy} shows typical instantiations: (i) \emph{Lexical} retrieval uses index statistics as priors, applies sparse matching as likelihood-like evidence, constrained by top-$k$ truncation; (ii) \emph{Dense} retrieval amortises document encoding as a prior-like representation, applies query--document similarity as evidence, constrained by ANN recall; (iii) \emph{Multi-stage} pipelines compose a cheap first-stage prior with a heavier re-ranker likelihood over a restricted candidate set; (iv) \emph{RAG} introduces an explicit context-budget operator that truncates which evidence can be presented to the generator. These instantiations motivate analysing retrieval failures as \emph{support restriction} failures: relevant documents may not be surfaced before expensive interaction can act.

\para{Interaction over Structured Spaces}

Query--document interaction rarely occurs over the full corpus. Instead, it is almost always applied to a candidate set produced by earlier retrieval stages~\cite{cambazoglu:2016}. As a result, interaction operates over a document space already structured by retrieval heuristics, representations, or prior rankings. This structure may be weak, as in BM25-based retrieval where only limited prior assumptions such as length normalisation are enforced~\cite{robertson:1995}, but it nonetheless constrains which documents interaction can affect. Viewing interaction as operating over a structured space helps explain why retrieval pipelines with very different architectures can exhibit similar behaviour. In the taxonomy (Table~\ref{tab:systems_taxonomy}), this appears as an explicit approximation operator: early-stage retrieval defines the support on which later likelihood refinement can act.

\para{The Purpose of Interaction}

Across retrieval systems, query--document interaction functions to filter candidate documents, reweight their relative importance, or gate access to downstream processing~\cite{wang:2011, cambazoglu:2016}. These three roles correspond to common inference patterns: filtering instantiates support restriction, reweighting instantiates score-level posterior reshaping on a fixed support, and gating instantiates budgeted access to subsequent computation. Interaction thus selects among pre-existing document distinctions rather than constructing relevance from scratch. This perspective helps explain why relatively simple fusion methods or shallow interaction models can perform competitively when combined with strong document-side signals~\cite{cormack:2009}.

\para{Beyond Ranking}

Ranking is only one stage within a broader retrieval pipeline that may include readers, generators, or iterative retrieval components~\cite{lewis:2020, pradeep:2021}. The taxonomy (Table~\ref{tab:systems_taxonomy}) highlights that this shift changes the approximation boundary, from ranking over candidates to selecting evidence under a hard context budget. In such regimes, query-independent signals (e.g., trust, redundancy, or generic utility) become first-class because they determine which evidence remains available to downstream reasoning under truncation. In retrieval-augmented generation systems, for example, ranking may primarily serve to surface plausible contexts rather than to fully resolve relevance~\cite{shi:2023, liu:2024}. The division of labour between ranking and downstream processing thus becomes an explicit design choice.

\begin{table}[t]
\caption{Systems as approximate inference, expressed per regime as prior, likelihood, and approximation operator.}
\label{tab:systems_taxonomy}
\centering
\small
\setlength{\tabcolsep}{2pt}
\renewcommand{\arraystretch}{1.15}
\begin{tabular}{@{}l p{0.77\columnwidth}@{}}
\toprule
Regime & Posterior view (prior / likelihood / approximation) \\
\midrule
Lexical &
\textbf{Prior}: index statistics, normalisation.\newline
\textbf{Likelihood}: lexical evidence aggregation.\newline
\textbf{Approx}: model-form bias over full-corpus scoring. \\[2pt]
\midrule
Dense &
\textbf{Prior}: cached embeddings (implicit).\newline
\textbf{Likelihood}: vector similarity (amortised).\newline
\textbf{Approx}: ANN search; collapsed prior$\times$likelihood. \\[2pt]
\midrule
Multi-stage &
\textbf{Prior}: early-stage scores / cached signals.\newline
\textbf{Likelihood}: expensive interaction / reranking.\newline
\textbf{Approx}: candidate truncation; staged refinement. \\[2pt]
\midrule
RAG  &
\textbf{Prior}: cached utilities, trust signals.\newline
\textbf{Likelihood}: retrieval as evidence proposal.\newline
\textbf{Approx}: context-budget truncation; evidence selection. \\
\bottomrule
\end{tabular}
\end{table}

\para{Queries as Constraints}
Queries can be interpreted as imposing constraints over a pre-computed document space rather than generating relevance independently. Under this view, query--document interaction restricts attention to documents whose attributes, content, or utility align with the expressed constraints~\cite{belkin:1980, ingwersen:1996}. This interpretation aligns with settings in which queries are short or underspecified, yet retrieval remains effective.

As offline document modelling becomes richer, more aspects of relevance can be resolved before query time~\cite{chang:2024, nogueira:2019b}. This does not eliminate the need for query--document interaction, but it changes its role and emphasis. Interaction may shift toward selecting, combining, or constraining pre-computed signals. This redistribution has implications for where learning effort is invested within a retrieval system. In taxonomy terms (Table~\ref{tab:systems_taxonomy}), strengthening offline modelling shifts systems along the prior axis and can convert unconstrained interaction into constrained operators such as selection, fusion, and budgeted refinement.

\section{Preliminary Findings}
In this section, we provide evidence towards answering the question:
``\emph{Which effectiveness gains are achievable through
document-side evidence, versus requiring query-conditioned
interaction?}'' --- one of our open questions posed in Section \ref{sec:application}.

\subsection{Rank Fusion with a Learned Prior}
We define a query-independent \emph{document utility} $u_\phi(d)\in\mathbb{R}$, interpreted as a log prior potential,
\begin{equation}
u_\phi(d)\;\approx\;\log \tilde p_\phi(d),
\label{eq:prior}
\end{equation}
where $\tilde p_\phi(d)$ is an unnormalised prior over documents. Utility is computed offline and cached for all $d\in\mathcal{D}$, enabling reuse with minimal additional document-side inference at query time (only the deployed scorer $s_\theta$ runs online with the addition of prior scores).

\para{Fusion in Log-Space}
\label{sec:addition}
Motivated by Eq.~\eqref{eq:bayes}, we combine the query-conditioned score and static utility via a fused potential
\begin{equation}
S(q,d) = F\bigl(s_\theta(q,d),\, u_\phi(d)\bigr),
\qquad d \in \mathcal{C}_q,
\label{eq:fused_score}
\end{equation}
chosen so that $S$ is an unnormalised log-posterior up to a query-dependent constant. We adopt additive fusion,
\begin{equation}
S_{\text{add}}(q,d) \;=\; s_\theta(q,d) + \gamma\, u_\phi(d),
\qquad \gamma \ge 0,
\label{eq:additive}
\end{equation}
which recovers the log-posterior of Eq.~\eqref{eq:bayes} when $s_\theta$ is calibrated as a log-likelihood and $u_\phi$ as a log-prior, and more generally defines a log-linear combination of potentials whose ranking depends only on $\gamma$ and the relative scale of the two scores.

Although additive score combination is well studied in learning-to-rank~\cite{fox:1993, liu:2009}, our setting differs in two respects. First, $u_\phi(d)$  and $s_\theta(q,d)$ are derived from the same modality yet are functionally orthogonal: the utility model sees strictly less input than the relevance scorer, so gains cannot stem from complementary data access but instead reflect document-level regularities that $s_\theta$ can represent in principle but struggles to disentangle from query-dependent signal during joint optimisation. Second, the prior is computed offline and cached, contributing zero marginal query-time cost and can be updated, audited, or replaced independently of the deployed model.

\subsection{Experimental Setup}

\para{Datasets}
We assess effectiveness on the TREC Deep Learning 2019~\citep{craswell:2019} and 2020~\citep{craswell:2020} test collections, retrieving from the MSMARCO passage collection~\citep{nguyen:2016}. All reported test collections apply normalised discounted cumulative gain (nDCG)~\citep{jarvelin:2002} as their primary measure, and we report this value at a rank cutoff of 10 as is standard.

\para{Inference Setting}
Our core setting fuses a static utility model \(u_\phi(d)\) to complement an existing dynamic ranker \(s_\theta(q,d)\). In this initial investigation, \(\theta\) and \(\phi\) are independent (no awareness of each other or respective score modes).
This ensures that improvements reflect complementary utility estimation rather than changes to the dynamic ranker. In this study we deliberately operate in an \emph{uncalibrated composition} setting: we do not learn fusion weights or perform score calibration between $s_\theta$ and $u_\phi$, and we fix the fusion parameter ($\gamma=1.0$) to isolate whether the offline utility signal is complementary in principle. This choice also preserves the intended deployment constraint: $u_\phi(d)$ is a cache lookup and does not introduce any additional document-side model forward pass at query time beyond the already-deployed scorer producing $s_\theta$. For simplicity, we perform fusion on the top-1000 documents retrieved by the retriever, rather than combining retrieval results with the full-corpus prior ranking, while still re-ranking only the top-100 for parity with baseline approaches.

\para{Quality Estimators.}
We experiment with fine-tuned models trained by \citet{chang:2024}. All checkpoints are based on the T5 encoder–decoder architecture~\cite{raffel:2020}. These models are trained as quality classifiers using binary cross-entropy over human-annotated relevant texts from the MSMARCO training corpus~\cite{nguyen:2016} and hard negatives drawn uniformly from the top-$k$ documents retrieved by BM25 for each training query. Functionally this training is identical to that of MonoT5~\cite{nogueira:2020} but without providing the query within the input context of the model. These models output log-odds of a document being of high quality. We run inference for all models on a single NVIDIA 4090 GPU with 24GB VRAM. This places reproducibility of our work within the hardware constraints of most academic labs. All experiments are carried out using the PyTerrier framework~\cite{pyterrier:2020} and its associated plug-ins. We release our codebase to assist reproducibility of our work and foster further research\footnote{\href{https://zenodo.org/records/18630068}{https://zenodo.org/records/18630068}}.

\para{Retrieval Models}
We employ two sparse first-stage retrieval models to generate candidate document sets for downstream scoring and analysis.

\uls
    \item \textbf{BM25}~\cite{robertson:1995}: A classical probabilistic term-weighting model that remains ubiquitous in modern search systems due to its simplicity, efficiency, and strong empirical performance. BM25 ranks documents by combining term frequency saturation and inverse document frequency, providing a strong lexical baseline. We use Robust04-tuned parameters (default in our chosen framework) being $k_1=1.2, b=0.75, k_3=8$.
    \item \textbf{SPLADE}~\cite{formal:2021}: A neural sparse retrieval model based on a BERT-style encoder that learns to expand and reweight query and document representations in the lexical space\footnote{We use the checkpoint trained through distillation naver/splade-cocondenser-ensembledistil~\cite{formal:2022}}. 
\ule

\para{Re-Ranking Models}
We re-rank our combined static/dynamic retrieval systems with two state-of-the-art re-rankers. In all cases, we re-rank the top 100 documents returned by a standalone or utility-fused retriever. 

\uls
    \item \textbf{MonoELECTRA (CE)}~\cite{schlatt:2024b}: An ELECTRA-based~\cite{clark:2020} point-wise cross-encoder trained using distillation from a larger list-wise ranking model. 
    \item \textbf{RankZephyr (LLM)}~\cite{pradeep:2023}: A Zephyr-based~\cite{tunstall:2023} list-wise cross-encoder trained using distillation from RankGPT~\cite{sun:2023}. We follow prior art, applying a window size of 20 documents with a stride of 10~\cite{pradeep:2023, parry:2024}.
\ule

\subsection{Results and Discussion}

When fusing BM25 with a learned document prior, observe in Table \ref{tab:msmarco}, that nDCG@10 significantly improves across all fusion settings. We observe a weak trend in effectiveness with the quality estimator's capacity, though this is primarily limited to QualT5-tiny, generally underperforming relative to larger variants. 
Under SPLADE, fusion consistently \emph{degrades first-stage} effectiveness, suggesting that the injected utility term is not complementary at the retrieval stage and may partially double-count signals already expressed by SPLADE's learned sparse expansion.
Notably, this does not preclude gains after re-ranking: improvements can still arise when downstream cross-encoders or LLM rerankers correct first-stage misorderings under restricted depth. This further motivates the trade-off between online and offline model capacity. We additionally note that in all but one re-ranking settings, fusion outperforms index pruning suggesting that priors can surface more valuable documents within the top-$k$ beyond reducing the ranking of low quality texts (similar to pruning).  

Some gains are observed with cross-encoder re-ranking, but larger gains are observed with the LLM re-ranker, which exhibits positional bias~\cite{schlatt:2024, parry:2024}. This point holds under both first-stages. We consider that a retriever may provide an out-of-distribution permutation for an LLM re-ranker and that in adding quality signals orthogonal to the retriever we may provide a better ordering for the LLM. Such a finding is particularly interesting in the case of RAG where a generator must choose context similarly to the list-wise ranker in order to generate output, an interesting future direction would be the direct optimisation of a prior to re-order candidates conditional on RAG effectiveness, crucially without having to consider a query.

\begin{figure*}[t]
    \centering
    \begin{subfigure}[t]{\columnwidth}
        \includegraphics[width=\linewidth]{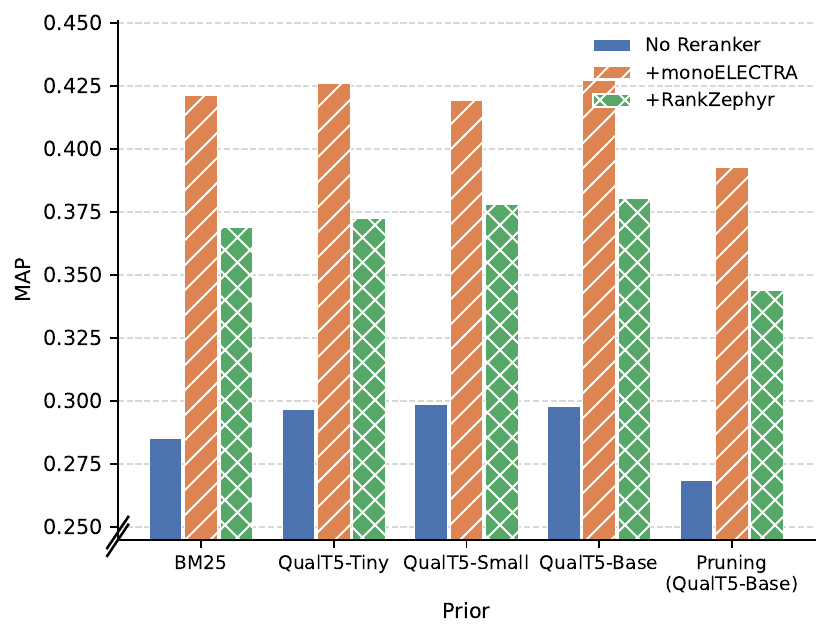}
        \caption{MAP}
    \end{subfigure}
    \hfill
    \begin{subfigure}[t]{\columnwidth}
        \includegraphics[width=\linewidth]{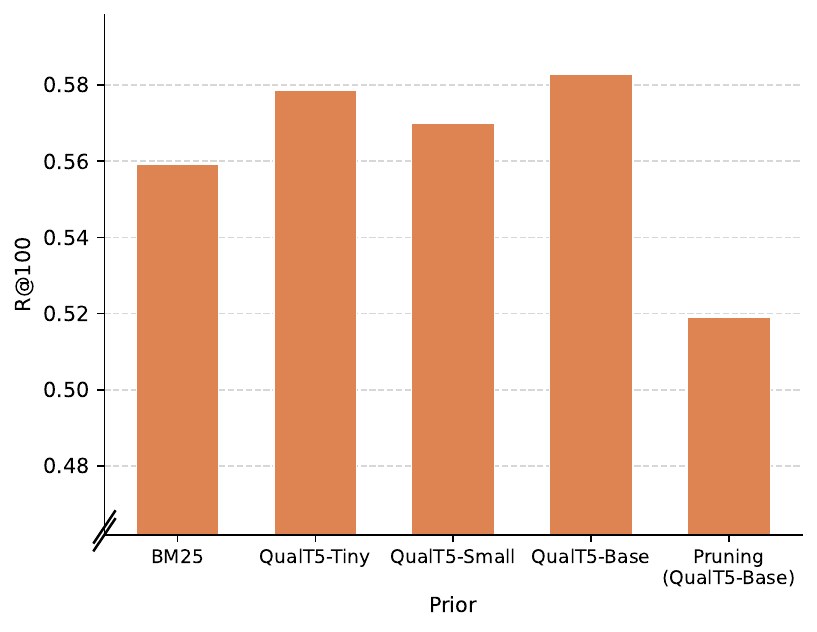}
        \caption{R@100}
    \end{subfigure}
    \caption{MAP and R@100 on TREC DL-2020 for BM25 under quality priors (QualT5-Tiny/Small/Base fusion and pruning, with and without reranking.}
    \label{fig:map-recall-dl20}
\end{figure*}

Considering deeper measures, under BM25, fusion improves both AP and 
R@100 as shown in Figure \ref{fig:map-recall-dl20}, indicating that the prior promotes relevant documents into the re-ranking window that lexical matching alone misses. Under SPLADE, both measures degrade marginally at the first stage, consistent with signal overlap. Crucially, fusion preserves R@1000 
(the candidate pool is reordered, not reduced), whereas pruning removes documents and reduces R@1000 by 3--9 absolute points. After re-ranking, AP gains are also present under SPLADE: RankZephyr on DL-2020 improves AP by approximately 0.03 despite first-stage degradation, suggesting that the prior produces candidate orderings which are favourable under the LLM ranker. Crucially, these models are not calibrated to a particular likelihood; this initial evidence points towards the potential for these priors to be used not only for efficiency through pruning but to concretely provide new signals in ranking to improve effectiveness. The appeal of the uncalibrated setting is that we can be sure that such a prior can still be used for pruning, an interesting problem remains in optimising a prior to serve both purposes.

\begin{table*}[ht!]
\centering
\caption{
Results retrieved over MSMARCO Passage TREC Deep Learning Collections. Significance is reported via a paired t-test ($\alpha=0.05$) with Bonferroni correction via $\dagger$. QualT5 models are trained with a binary cross-entropy loss. Pruning is applied to remove the bottom $20^{\text{th}}$ percentile of documents by estimated quality. In all cases, fusion is an additive fusion with equal weighting (See Section \ref{sec:addition}). Overall values are computed via geometric mean.
\label{tab:msmarco}
}
\small
\begin{tabular}{ll rrr rrr rrr}
\toprule
&
& \multicolumn{3}{c}{TREC DL-2019}
& \multicolumn{3}{c}{TREC DL-2020}
& \multicolumn{3}{c}{Overall} \\
\cmidrule(lr){3-5}
\cmidrule(lr){6-8}
\cmidrule(lr){9-11}
Type & Pipeline
& $1^{\text{st}}$ Stage & +CE & +LLM
& $1^{\text{st}}$ Stage & +CE & +LLM
& $1^{\text{st}}$ Stage & +CE & +LLM \\

\midrule
\multirow{2}{*}{Baseline} & BM25 & .4737$^{\phantom{\dagger}}$ & .7132$^{\phantom{\dagger}}$ & .6604$^{\phantom{\dagger}}$ & .4712$^{\phantom{\dagger}}$ & .6986$^{\phantom{\dagger}}$ & .6233$^{\phantom{\dagger}}$ & .4724$^{\phantom{\dagger}}$ & .7059$^{\phantom{\dagger}}$ & .6419$^{\phantom{\dagger}}$ \\
 & BM25 + Pruning (QualT5~\cite{chang:2024}) & .4947$^{\phantom{\dagger}}$ & \textbf{.7283}$^{\phantom{\dagger}}$ & .6549$^{\phantom{\dagger}}$ & .4631$^{\phantom{\dagger}}$ & .6771$^{\phantom{\dagger}}$ & .6280$^{\phantom{\dagger}}$ & .4789$^{\phantom{\dagger}}$ & .7027$^{\phantom{\dagger}}$ & .6415$^{\phantom{\dagger}}$ \\
\cmidrule(r){2-11}
\multirow{3}{*}{Fusion} & + QualT5-Tiny & .5180$^\dagger$ & .7159$^{\phantom{\dagger}}$ & .6709$^{\phantom{\dagger}}$ & .4931$^{\phantom{\dagger}}$ & .7036$^{\phantom{\dagger}}$ & .6506$^{\phantom{\dagger}}$ & .5056$^{\phantom{\dagger}}$ & .7097$^{\phantom{\dagger}}$ & .6607$^{\phantom{\dagger}}$ \\
 & + QualT5-Small & \textbf{.5195}$^\dagger$ & .7155$^{\phantom{\dagger}}$ & \textbf{.6767}$^{\phantom{\dagger}}$ & \textbf{.5002}$^\dagger$ & \textbf{.7057}$^{\phantom{\dagger}}$ & \textbf{.6774}$^\dagger$ & \textbf{.5099}$^{\phantom{\dagger}}$ & .7106$^{\phantom{\dagger}}$ & \textbf{.6770}$^{\phantom{\dagger}}$ \\
 & + QualT5-Base & .5142$^\dagger$ & .7252$^{\phantom{\dagger}}$ & .6762$^{\phantom{\dagger}}$ & .4925$^{\phantom{\dagger}}$ & .6980$^{\phantom{\dagger}}$ & .6457$^{\phantom{\dagger}}$ & .5034$^{\phantom{\dagger}}$ & \textbf{.7116}$^{\phantom{\dagger}}$ & .6610$^{\phantom{\dagger}}$ \\
\midrule
\multirow{2}{*}{Baseline} & SPLADE & \textbf{.7312}$^{\phantom{\dagger}}$ & .7685$^{\phantom{\dagger}}$ & .6789$^{\phantom{\dagger}}$ & \textbf{.7204}$^{\phantom{\dagger}}$ & \textbf{.7686}$^{\phantom{\dagger}}$ & .6669$^{\phantom{\dagger}}$ & \textbf{.7258}$^{\phantom{\dagger}}$ & \textbf{.7685}$^{\phantom{\dagger}}$ & .6729$^{\phantom{\dagger}}$ \\
 & SPLADE + Pruning (QualT5~\cite{chang:2024}) & .7256$^{\phantom{\dagger}}$ & .7573$^{\phantom{\dagger}}$ & .7175$^{\phantom{\dagger}}$ & .6995$^{\phantom{\dagger}}$ & .7489$^{\phantom{\dagger}}$ & .6692$^{\phantom{\dagger}}$ & .7126$^{\phantom{\dagger}}$ & .7531$^{\phantom{\dagger}}$ & .6934$^{\phantom{\dagger}}$ \\
\cmidrule(r){2-11}
\multirow{3}{*}{Fusion} & + QualT5-Tiny & .6983$^\dagger$ & .7686$^{\phantom{\dagger}}$ & .7285$^{\phantom{\dagger}}$ & .6391$^\dagger$ & .7657$^{\phantom{\dagger}}$ & \textbf{.7143}$^{\phantom{\dagger}}$ & .6687$^{\phantom{\dagger}}$ & .7672$^{\phantom{\dagger}}$ & .7214$^{\phantom{\dagger}}$ \\
 & + QualT5-Small & .7142$^\dagger$ & \textbf{.7694}$^{\phantom{\dagger}}$ & \textbf{.7395}$^{\phantom{\dagger}}$ & .6706$^\dagger$ & .7644$^{\phantom{\dagger}}$ & .7132$^{\phantom{\dagger}}$ & .6924$^{\phantom{\dagger}}$ & .7669$^{\phantom{\dagger}}$ & \textbf{.7264}$^{\phantom{\dagger}}$ \\
 & + QualT5-Base & .7141$^{\phantom{\dagger}}$ & .7661$^{\phantom{\dagger}}$ & .7154$^{\phantom{\dagger}}$ & .6502$^\dagger$ & .7634$^{\phantom{\dagger}}$ & .6741$^{\phantom{\dagger}}$ & .6821$^{\phantom{\dagger}}$ & .7647$^{\phantom{\dagger}}$ & .6947$^{\phantom{\dagger}}$ \\
\bottomrule
\end{tabular}
\end{table*}

\para{Qualitative Differences}
In investigating rankings produced by re-ranking pipelines with and without QualT5 fusion, we observe a general trend: adding the prior largely improves navigational queries. Intents associated with definitions, for instance, biographical queries or definitions of words, or exact quantifications, such as the cost of an entity or when an event occurred, tended to improve under both monoELECTRA and RankZephyr re-ranking. A plausible explanation is that a query-independent quality prior disproportionately helps intents where relevance is strongly mediated by document-side ``usability'' (e.g., concise definitional pages, canonical entity descriptions, or well-formed fact statements), because these properties are stable across queries and therefore well matched to offline estimation. By contrast, open-ended exploratory queries often require query-specific evidence aggregation across multiple plausible facets; in these cases, injecting a global utility term can over-prefer generally well-written or authoritative documents that are only weakly aligned with the particular causal or explanatory frame implied by the query, yielding small degradations that downstream interaction must correct.

\section{Implications and Applications}
\label{sec:application}

\begin{figure}
    \centering
    \includegraphics[width=\linewidth]{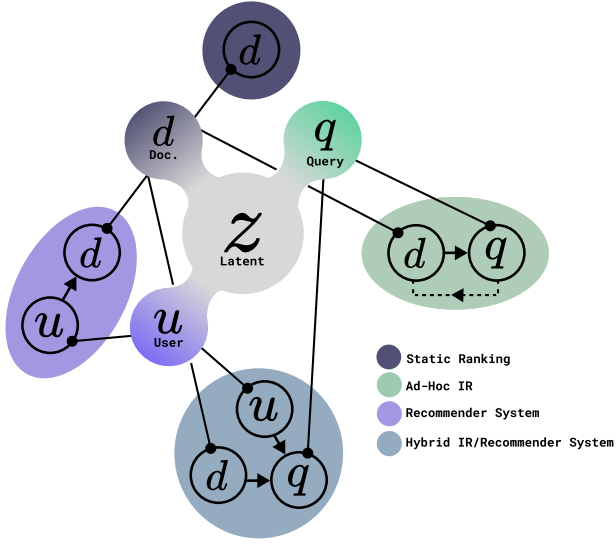}
    \caption{An illustration of how posterior inference of $p(d|q)$ allows us to connect not only different aspects of information access but also express ad-hoc pipelines as formal objects. Re-ranking is seen as an iterative refinement of likelihood, here expressed as taking the outcome of some first stage, which governs our second stage.}
    \Description{A diagram showing four retrieval regimes arranged around a central latent variable z. The top-left shows a document node d alone (static ranking). The top-right shows d connected to q (ad-hoc IR). The bottom-left shows d connected to u (recommender system). The bottom-right shows d connected to both q and u (hybrid IR/recommender system). A legend distinguishes the four existing regimes by colour.}
    \label{fig:layers}
\end{figure}

Framing retrieval as approximate posterior inference turns several ``systems'' questions into modelling questions: (i) which computation should be amortised offline, (ii) which must remain online to resolve intent, and (iii) how should multiple sources of belief be composed. Figure~\ref{fig:layers} illustrates this perspective: retrieval regimes differ not in objective but in which variables ($d$, $q$, $u$, $z$) are retained or marginalised, and in how prior and likelihood components are composed. We organise implications around three roles that priors play in a pipeline: \emph{efficiency}, \emph{effectiveness}, and \emph{control}. Finally we make a broader connection to recommender systems as hybrid information access becomes more prevalent.

\subsection{Priors for Efficiency}
Under the posterior view, efficiency is a consequence of shaping the posterior support so that expensive online inference is applied to fewer candidates.

\para{Index-side filtering.}
The prior can reduce the candidate pool before any interaction, either by permanently removing low-utility documents (static pruning) or by dynamically gating eligibility for later stages. This is not an auxiliary heuristic: it is an approximation to posterior inference in which documents with low prior mass are excluded before evaluating the likelihood. First-stage retrieval and re-ranking depth therefore become \emph{joint} design variables: an informative $u(d)$ permits reducing $k$ while preserving posterior mass on relevant documents.

\para{Context-side filtering.}
When the downstream component is an LLM, the dominant cost shifts to context construction and generation. The same prior mechanism acts as a context filter, determining which passages are admissible evidence and how much evidence is necessary, by preferring documents whose stable properties (reliability, coherence, coverage) make them useful across many queries.

\para{Trading prior strength for online capacity.}
If priors absorb stable document-side structure, the online model can be smaller while maintaining comparable posterior quality. Conversely, given a fixed online model, stronger priors can reduce re-ranking depth or context budget. This makes explicit a design trade-off usually treated implicitly through engineering constraints.

\para{Open Problems}
We encourage the community to investigate the following open problems related to the topic of how the document priors may be utilised for increasing the efficiency of rankers.
\begin{itemize}
    \item The amortised inference framing (Section~\ref{sec:bridge}) implies that prior cost should be paid once and amortised over the query workload: how should prior capacity scale with the cost profile of the downstream stage so that this trade-off is favourable across cross-encoder, LLM reranker, and RAG methodologies?
    \item Since first-stage depth and re-ranking depth are joint design variables under an informative prior, how should candidate set size and re-ranking depth be jointly optimised with prior strength under a fixed latency or token budget?
    \item How should priors be updated under corpus drift without retraining query-time models, given that the decomposition is intended to decouple offline and online components?
\end{itemize}

\subsection{Priors for Effectiveness}
The decomposition clarifies two distinct mechanisms. The first is \emph{additive}: priors inject stable evidence that the likelihood is not expected to infer. The second is \emph{corrective}: priors compensate systematic error induced by the deployed likelihood.

\para{Adding stable evidence.}
Document priors can encode properties that are query-invariant and difficult to infer from interaction alone: source reliability, structural completeness, writing quality, or domain compliance. Under the posterior view, these are not reranking tricks but explicit factors of posterior belief. This becomes salient in task-augmented settings where documents must be not only topically relevant but \emph{usable} as evidence (e.g., answerability, extractability, citation-worthiness), enabling the likelihood to focus on intent resolution.

\para{Correcting systematic error.}
End-to-end rankers absorb biases from supervision, negative sampling, and candidate generation; term-weighting models are restricted to exact matching. Many of these biases are stable across queries (popularity effects, domain over-exposure, annotation artefacts) and appear as persistent residual error in $s_\theta(q,d)$. Training a prior against a frozen likelihood provides a correction term that re-centres posterior scores where the likelihood is consistently miscalibrated, without modifying the online model. This decomposition also supports clearer \emph{attribution}: query-side failures (intent, ambiguity) and document-side failures (quality, trust, distributional skew) can be diagnosed independently rather than conflated inside a monolithic scorer.

\para{Open Problems} The following is a list of open problems, we suggest, that could be investigated to improve retrieval effectiveness with document priors.
\begin{itemize}
    \item Our preliminary findings show gains concentrated on navigational and definitional queries where document-side usability is stable: which effectiveness gains are primarily achievable through document-side evidence more broadly, versus requiring query-conditioned interaction?
    \item The corrective mechanism relies on biases being stable across queries; how can priors correct known biases of a deployed likelihood (e.g., popularity, recency, domain imbalance) without inducing new pathologies when that stability assumption breaks down?
    \item Can priors trained on one corpus transfer to correct likelihood-specific residuals on another, given that the prior is by design independent of the query-time model?
\end{itemize}

\subsection{Priors as Control Surfaces}
\label{sec:control}
Making priors explicit exposes controllable degrees of freedom typically entangled within end-to-end training. Control corresponds to adjusting prior terms while holding the likelihood fixed, changing posterior behaviour without changing query understanding.

\para{Global control.}
Safety, trustworthiness, diversity, and exposure policies are commonly introduced via manual boosts or filters. Under the posterior framing, these are prior terms that can be learned, calibrated, and audited independently of the likelihood. Policy updates can be deployed by modifying priors alone, and audits can target the prior directly.

\para{User control.}
Personalisation fits naturally as a user-conditioned prior $p(d \mid u)$ that interacts with $s_\theta(q,d)$, matching the recommender-systems view of user--item affinity while preserving a distinct role for the likelihood to resolve intent. The key question is whether personalisation can be controlled primarily through priors, leaving the likelihood as a generic intent model.

\para{Composing multiple priors.}
A single posterior can combine modular utilities,
\begin{equation}
\log p(d \mid q) \;\propto\; s_\theta(q,d) \;+\; \sum_{m} \lambda_m\, u_m(d),
\end{equation}
where each $u_m$ encodes a distinct belief (quality, trust, freshness, user affinity). Because these sources differ in supervision, stability, and update frequency, they are difficult to express in a single end-to-end objective but natural to maintain independently as priors composed with a shared likelihood. Improvements to any $u_m$ propagate across all pipelines sharing a corpus without retraining the online model, directly reducing the coupling that makes end-to-end system iteration brittle.

\para{Open Problems}
\begin{itemize}
    \item Since control is achieved by modifying priors while holding the likelihood fixed, how should propensity and exposure constraints be expressed as prior terms and tuned against relevance evidence without destabilising the decomposition?
    \item The modular composition above assumes priors are independently valid; what monitoring and auditing methods are needed to validate each prior independently of the query model (e.g., drift detection, calibration checks, policy compliance)?
\end{itemize}

\subsection{Connecting to Broader Information Access}
Conditioning priors on richer context, including users, domains, tasks, or temporal signals, extends the control surface beyond global document properties. Each additional conditioning variable introduces a separable degree of freedom that can be tuned, audited, or removed without modifying the query-time likelihood. This generalisation connects the posterior view to settings beyond ad-hoc retrieval, most directly to recommender systems.

\para{Connection to Recommender Systems.}
The prior--likelihood decomposition bridges ad-hoc retrieval and recommendation, two settings that share a common inferential structure despite being treated as distinct research areas. As Figure~\ref{fig:layers} illustrates, the key difference is which variables are retained: ad-hoc IR conditions on $(d, q)$, recommendation on $(d, u)$, and hybrid systems on $(d, q, u)$ with static ranking as the limiting case where only $d$ is modelled. The latent variable $z$ mediates between these regimes, representing corpus structure that may be marginalised or explicitly resolved depending on the system.

In collaborative filtering, a user--item score is decomposed as $\hat{r}(u,i) = f(\mathbf{p}_u, \mathbf{q}_i)$, where $\mathbf{q}_i$ encodes item properties and $\mathbf{p}_u$ encodes user preferences, learned jointly from interaction histories~\cite{koren:2009, he:2017}. Under our framework, $\mathbf{q}_i$ plays the role of a user-marginalised item prior while $\mathbf{p}_u$ conditions this prior on a specific user, analogous to how a query conditions a document prior on a specific need. Both settings face the same allocation problem: which aspects of the score can be precomputed from one side (documents/items) and which require observing the other (queries/users) at inference time.

This becomes concrete in hybrid IR/recommendation settings (news recommendation, e-commerce search, conversational assistants) that combine queries with user profiles. Personalised retrieval systems leverage persistent user profiles, including interaction history, topical preferences, and reading level, as priors that modulate relevance estimates~\cite{salemi:2024, pasi:2018, bassani:2024}. Formally, such systems admit a three-way factorisation,
\begin{equation}
  p(d \mid q, u)
    \;\propto\;
    p(q \mid d)\;
    p(d \mid u)\;
    p(d),
  \label{eq:hybrid}
\end{equation}
where each term operates on a different timescale: corpus updates for $p(d)$, user history evolution for $p(d \mid u)$, and per-request intent for $p(q \mid d)$. Only the query likelihood must be evaluated under strict latency constraints. The decomposition also unifies both communities' approaches to cold-start: in recommendation (new users/items lacking history) and in retrieval (underspecified queries or novel documents), the remedy amounts to increasing prior weight when the likelihood signal is weak. Exposure and propensity constraints, routine in recommendation, are naturally expressible as prior adjustments in retrieval, counteracting popularity bias without modifying the query-conditional ranker.

\para{Open Problems}
\begin{itemize}
    \item The three-way factorisation in Equation~\ref{eq:hybrid} separates user-conditioned priors from query likelihoods; can priors learned from recommendation interaction logs transfer to ad-hoc retrieval to improve personalisation without sacrificing topical precision?
    \item Under the cold-start unification above, how should prior--likelihood weighting adapt as user history length varies, and what supervision strategies are needed when retrieval queries are one-shot rather than repeated?
    \item What is the relationship between retrievability-based document priors~\cite{azzopardi:2008} and item popularity priors in recommendation, and can a unified formulation serve both settings?
\end{itemize}

\section{Conclusion}
We framed retrieval pipelines as approximate inference procedures: systems combine reusable, query-independent beliefs about documents (priors) with query-conditioned evidence (likelihood-like scoring) under explicit approximation operators such as top-$k$ truncation, ANN recall limits, and context budgets. This view makes a concrete design question explicit: which beliefs should be amortised offline, and how should they be composed with online scoring under fixed compute constraints. Our preliminary findings suggest that a learned document-utility prior can be complementary to classical first-stage retrievers and can yield larger gains for LLM-based re-ranking, while naive fusion can degrade first-stage performance for strong learned sparse retrievers, consistent with overlap or miscalibration between signals. Overall, the posterior lens motivates treating priors as first-class system components that can be computed and updated offline, and reused across rankers. These proposed areas of improvement, efficiency, effectiveness and control, are closely aligned with long-standing community concerns, yet have been comparatively under-explored as modelling effort has shifted toward ever more expensive query-time relevance estimation; making priors explicit refocuses attention on decomposed system design and renders the resulting support--quality--cost trade-offs measurable.

\bibliographystyle{ACM-Reference-Format}
\balance
\bibliography{refs}
\end{document}